\documentclass[fleqn,twoside]{article}
\usepackage{espcrc2}

\usepackage{epsfig}

\newcommand{\eq}{\begin{equation}}
\newcommand{\eqx}{\end{equation}}
\newcommand{\eqn}{\begin{eqnarray}}
\newcommand{\eqnx}{\end{eqnarray}}
\newcommand{\nc}{$ N_{\rm cut\ } $}

\def\Journal#1#2#3#4{{#1} {\bf #2} (#4) #3}
\def\NPB{{\em Nucl. Phys.} B}

\def\PRD{{\em Phys. Rev.} D}

\def\IJMPA{{\em Int. J. Mod. Phys.} A}

\def\be{\begin{equation}}
\def\ee{\end{equation}}

\newcommand{\AmS}{{\protect\the\textfont2
  A\kern-.1667em\lower.5ex\hbox{M}\kern-.125emS}}

\begin{document}
\centerline{\LARGE\bf Hamiltonian study of Supersymmetric}
\vspace*{0.5cm}
\centerline{\LARGE\bf  Yang-Mills Quantum Mechanics$^{*+}$}
\vspace*{1.5cm}
\centerline{\Large J. KOTANSKI  and  J. WOSIEK}
\begin{center}
M. Smoluchowski Institute of Physics,
 Jagellonian University,\\
Reymonta 4, 30-059 Cracow, Poland
\end{center}
\vspace*{2cm}

\centerline{\Large\bf Abstract}
\vspace*{0.5cm}
\begin{center}
\begin{minipage}{12cm}
{\large
New results obtained within the recently developed approach to
 supersymmetric quantum mechanical systems are presented. The method does not suffer from
the sign problem
in any dimensions and is capable to provide any quantum observable with controllable systematic
error. Discussed results include: the spectrum and Witten index of the D=4 system,
and the spectrum of zero volume glueballs in higher $( 4\le D \le 10 )$ dimensions.
}
\end{minipage}
\end{center}

\vspace*{1cm}
\vspace*{2cm}
\newline \noindent TPJU-16/02
\newline August 2002 \newline

\vspace*{\fill}
----------------------------------\newline
{\small $^*$ Presented by J. Wosiek on the XX Symposium on Lattice
Field Theory, MIT, Cambridge MA, June 2002.\newline $^+$ Supported
by the Polish Committee for Scientific Research under the grant 2
P03B 09 622 during 2002 -- 2004.}\newline
\newpage
%
\title{Hamiltonian study of Supersymmetric Yang-Mills Quantum Mechanics\thanks{Supported by the
Polish Committee for Scientific Research under the grant 2 P03B 09
622 during 2002 -- 2004.}
\thanks{Presented by J. Wosiek.}
}

\author{J. Kotanski  and  J. Wosiek \address{M. Smoluchowski Institute of Physics,
 Jagellonian University,\\ Reymonta 4,
30-059 Cracow, Poland }  }

\begin{abstract}
New results obtained within the recently developed approach to
 supersymmetric quantum mechanical systems are presented. The method does not suffer from
the sign problem
in any dimensions and is capable to provide any quantum observable with controllable systematic
error. Discussed results include: the spectrum and Witten index of the D=4 system,
and the spectrum of zero volume glueballs in higher $( 4\le D \le 10 )$ dimensions.
\end{abstract}

\maketitle

We report on the new development in studying supersymmetric
quantum mechanical systems.
The main goal is to solve the ten (space-time) dimensional (D=10) supersymmetric Yang-Mills
quantum mechanics (SYMQM) with the SU(N) gauge group \cite{CH}. At infinite N it provides
 relatively simple model of M-theory\cite{BFSS}\footnote{For finite N it represents
 compactified version of
M-theory.}.
 In spite of recent results that question the exact equivalence between the two,
D=10 SYMQM possesses  lot of fascinating properties required in
M-theory  (e.g. threshold bound states). Independently, these
systems provide a good  laboratory to study various consequences
of supersymmetry.

     We use the  hamiltonian formalism adapted  to supersymmetric systems
with local gauge invariance \cite{JW}.
It consists of two steps: 1) creation of a finite basis of gauge invariant states,
and 2) algebraic calculation of the matrix representation of the hamiltonian,
 followed by its numerical diagonalization. Algebraic calculations are automated by
implementing standard rules of quantum mechanics in an symbolic language like Mathematica.
Faster, compiler based, version is now also available. Given the spectrum and energy
eigenstates, any other relevant observable can be readily obtained using our computer
based rules of quantum mechanics.

Such an approach necessarily introduces a cutoff - the size of the
basis. We choose for it the gauge and rotationally invariant total
number of bosonic quanta $B=a_a^i a_a^{i\dagger}$. Hence the cut
off basis consists of all states with $B \le N_{\rm cut} $ and all
allowed fermions. Changing the size of the basis gives the
quantitative measure of the finite \nc effects similarly to
lattice calculations.

 The technique has been applied to Wess-Zumino quantum mechanics,
D=2 and D=4 SYMQM and to D = 5 - 10 YMQM, all based on the SU(2) gauge group \cite{JW}.
In all cases studied until now the spectrum of lower states converges with \nc
well within the reach of present computers. Clear SUSY signatures are seen.
Similar methods have been independently
developed in \cite{JAP} for lower dimensional supersymmetric field theories\footnote{See
 Uwe Trittmann contribution.}. Since \cite{JW} a notable progress,
in increasing the Fock space, has been achieved
by M. Campostrini. Here we report the new results for four (space-time) dimensional
quantum mechanics.

\noindent{\em D=4 SYMQM.}\newline
The system is described by nine bosonic coordinates $ x^i_a(t) $,
$i=1,2,3; a=1,2,3$ and six independent fermionic ones contained in the Majorana spinor
$\psi_a^{\alpha}(t)$, $\alpha=1,...,4$.
Hamiltonian reads  \cite{CH}
\eqn
H&=&{1\over 2} p_a^ip_a^i+{g^2\over 4}\epsilon_{abc} \epsilon_{ade}x_b^i x_c^j x_d^i x_e^j\\ \nonumber
 & + & {i g \over 2} \epsilon_{abc}\psi_a^T\alpha^k\psi_b x_c^k ,  \label{HD4}
\eqnx
with the Dirac $\alpha$
 matrices.  We work in the Majorana representation of Itzykson and Zuber.

The system has rotational symmetry,
with the spin(3) angular momentum
$
J^i=\epsilon^{ijk}\left( x^j_a p^k_a-{1\over 4}\psi^T_a\Sigma^{jk}\psi_a\right),\;\;\;
\Sigma^{jk}=-{i\over 4}[\Gamma^j,\Gamma^k],
$
the gauge invariance with the generators
$
G_a=\epsilon_{abc}\left(x_b^k p_c^k-{i\over 2}\psi^T_b\psi_c \right), 
$
and supersymmetry generated by the charges
$
Q_{\alpha}=(\Gamma^k\psi_a)_{\alpha}p^k_a + i g \epsilon_{abc}(\Sigma^{jk}\psi_a)_{\alpha}x^j_b x^k_c.
$

Generation of basis, construction of Majorana fermions and details of the calculation
of hamiltonian matrix and other observables are described in \cite{JW}.
Here we present new results for higher cutoffs ( \nc $\leq 8 $ )
 in all fermionic sectors. Hamiltonian (\ref{HD4}) conserves
fermion number $F=f_a^i f_a^{i\dagger}$, hence it suffices to diagonalize H separately in each
fermionic sector. Moreover, as a consequence of a particle-hole symmetry
there are only four independent sectors.
\begin{figure}[h]
 \epsfxsize=17pc \epsfysize=14pc
 \epsfbox{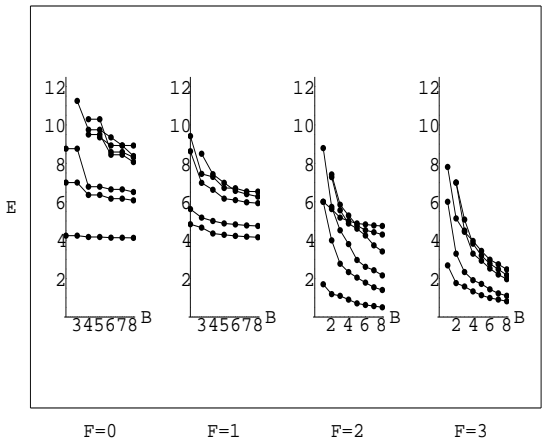}
 \caption{Cut-off dependence of the lower levels in all four fermionic sectors.}
\label{fig1}
\end{figure}

The spectrum of the theory (Fig.1) is rather rich and behaves differently in different sectors.
In $F=0$ sector our spectrum converges to the well known classic results of
L\"{u}scher and M\"{u}nster for zero volume glueballs \cite{LM} providing satisfactory
test of the whole approach\footnote{Together with van Baal we have also found complete agreement
(for $F=0\; {\rm and}\; F=2$) with the recent adaptation of his
 method of supersymmetric spherical harmonics \cite{VB}.}. For $F=1$ we again observe fast convergence with
increasing \nc --
our bases are adequate to extract the "infinite volume" limit of lower glueballs and
gluinoballs in both sectors. One also observes emergence of the supersymmetric multiplets
spanned across different fermionic sectors\footnote{SUSY multiplets can be identified in detail
using supersymmetric charges  and angular momentum operator.}. In "fermion rich"
sectors $F=2,3$ situation is even more interesting. In addition to quickly convergent with \nc
states we observe levels with strong \nc dependence whose energies tend to zero.
In fact the cutoff dependence can be used to identify discrete and continuous spectra.
It was found that localized, discrete bound states have fast (e.g. exponential) convergence
with \nc , while continuous, non localized spectrum is signaled by slow $O(1/N_{cut})$
dependence. Therefore our results confirm (and quantify) famous SUSY predictions of the
continuous spectrum in this model \cite{DWLN}. We also establish that the SUSY vacuum is
the lowest, $J=0$, state in the $F=2$ sector, that it belongs to the continuum hence
is non normalizable, and in fact is doubly degenerate ($F=2\; {\rm and}\; 4$  sectors).

\begin{figure}[h]
 \epsfxsize=17pc \epsfysize=12pc
 \epsfbox{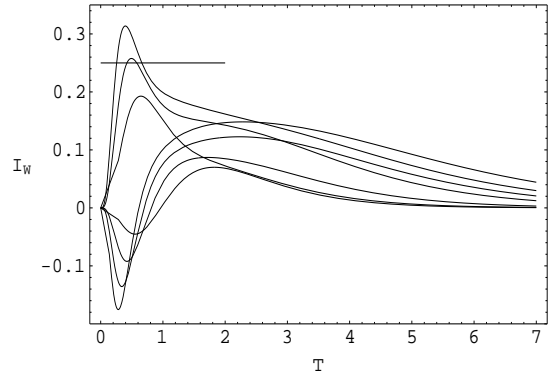}
 \caption{Witten index of D=4 SYMQM for up to 8 bosonic quanta. } \label{fig:toosmall}
\end{figure}
Witten index $ I_W(T)=\Sigma_{i} (-1)^{F_i}\exp{(-T E_i)}$, provides a global measure of the
restoration/violation of SUSY. It is discontinuous at $T=0$ with $I_W(0)$ depending on
the regularization. For discrete spectrum it is time independent and equals to the number of
ground states. For continuous spectrum it may have a mild $T$ dependence, still $I_W(\infty)$
gives the number of localized ground (unbalanced by SUSY) states. The index shown in Fig.2
was obtained directly from our spectrum.Compared to \cite{JW} we observe further slow approach
to the large \nc limit, consistent with all above properties. The physical (bulk) value
$\lim_{T\rightarrow 0} I_W(T)$
seems to point towards $1/4$ calculated from nonabelian integrals \cite{YI} and marked by the flat
line.
 There is no indication of the threshold bound state in agreement with Fig.1 and with theoretical
expectations.

\noindent{\em From four to ten space-time dimensions.}
To assess the feasibility of our approach to the full D=10 theory we have calculated the
spectra of the purely bosonic hamiltonians in all intermediate dimensions $ 4 \le D \le 10 $.
Thereby results shown in Fig.3 extend for the first time
the spectrum  of the zero volume glueballs to higher dimensions. This generalization is straightforward
in our approach, however it demands larger, but realistic, computer effort. In all dimensions
we see fast convergence with \nc indicating the localized nature of glueball bound states.
Since our cutoff respects rotational symmetry for all d=D-1, we observe exact SO(d) degeneracies
in the spectra. The ordering of the first three
multiplets, shown in Fig.3, remains the same as in $d=3$: scalar, two dimensional
symmetric tensor with dimension $g=d(d+1)/2-1$,
and scalar again. The largest uncertainty (i.e. that of the third d=9 level) is around 25\%
at present and diminishes gradually to below 1\% at $d=3$. Turning on d as another parameter
allows for further confrontation with analytical calculations. For example
the simple eye-ball estimate from Fig.3 suggest the power dependence  $ E \sim d^{\sim 1.5} $
to be compared with the mean field result at low temperature $d^{4/3}$ \cite{KAB}.

\begin{figure}[htb]
\vspace{9pt}
\epsfxsize=17pc \epsfbox{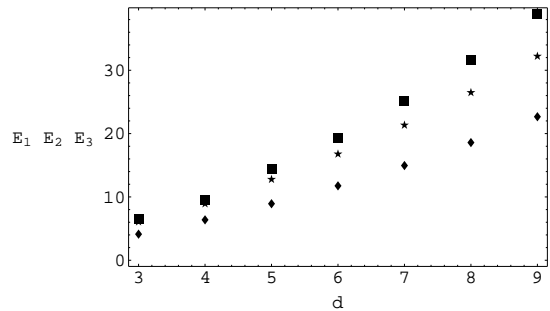} \caption{Zero volume glueballs in higher dimensions.} \label{fig3}
\end{figure}

We conclude that calculations in higher dimensions including $d=9$ are perfectly realistic
even if more time consuming. Consequently a path towards the quantitative study of the full
D=10 supersymmetric theory is now available.

\noindent{\em Acknowledgements.} I thank Theory Groups of Boston and Brown Universities,
especially C.-I Tan and R. C. Brower, for their hospitality.


\end{document}